\documentclass[a4paper,11pt]{article}
\pdfoutput=1 

\usepackage{jheppub} 

\usepackage{amsmath}
\usepackage{amsfonts,amssymb}

\usepackage[T1]{fontenc} 
\usepackage{bm}

\def\({\left(} 
\def\){\right)}

\usepackage{xcolor}
\xdefinecolor{redviolet}{RGB}{228,38,207}
\xdefinecolor{brickred}{RGB}{220,60,66}
\xdefinecolor{bluscuro}{RGB}{51,51,179}

\title{\boldmath Towards muon-electron scattering at NNLO}

\author[a]{Carlo M. Carloni Calame,}
\author[b]{Mauro Chiesa,}
\author[a]{Syed Mehedi Hasan,}
\author[c,a]{Guido Montagna,}
\author[a]{Oreste Nicrosini}
\author[a]{and Fulvio Piccinini}

\affiliation[a]{INFN, Sezione di Pavia, Via A. Bassi 6, 27100 Pavia, Italy}
\affiliation[b]{LAPTh, CNRS, Annecy, F-74940}
\affiliation[c]{Dipartimento di Fisica, Universit\`a di Pavia, Via A. Bassi 6, 27100 Pavia, Italy}

\emailAdd{carlo.carloni.calame@pv.infn.it}
\emailAdd{mauro.chiesa@lapth.cnrs.fr}
\emailAdd{syedmehe@pv.infn.it}
\emailAdd{guido.montagna@pv.infn.it}
\emailAdd{oreste.nicrosini@pv.infn.it}
\emailAdd{fulvio.piccinini@pv.infn.it}

\abstract{The recently proposed MUonE experiment at CERN aims at providing
  a novel determination of the leading order hadronic contribution to
  the muon anomalous magnetic moment through the study of elastic
  muon-electron scattering at relatively small momentum transfer. The 
  anticipated accuracy of the order of 10ppm demands for high-precision
  predictions, including all the relevant radiative corrections.
  The theoretical formulation for the fixed-order NNLO photonic 
  radiative corrections is described and the impact of the numerical
  results obtained with the corresponding Monte Carlo code is discussed
  for typical event selections of the MUonE experiment. 
  In particular, the gauge-invariant subsets of corrections due to
  electron radiation as well as to muon radiation are treated exactly. 
  The two-loop contribution due to diagrams where at least two virtual
  photons connect the electron and muon lines is approximated
  taking inspiration from the classical Yennie-Frautschi-Suura approach. 
  The calculation and its Monte Carlo implementation pave the 
  way towards the realization of a simulation code incorporating the full
  set of NNLO corrections matched to multiple photon radiation,
  that will be ultimately needed for data analysis.
}


\preprint{LAPTH-029/20}

\begin{document}
\maketitle
\flushbottom

\section{Introduction}
\label{sec:intro}

The value of the anomalous magnetic moment of the muon, $a_\mu = (g-2)_\mu / 2$, is a 
fundamental quantity in particle physics, that is presently known with 
a relative accuracy of 0.54ppm~\cite{Bennett:2006fi}. The muon anomaly, {\it i.e.} the deviation of the
magnetic moment from the value predicted by Dirac theory, is due to quantum loop corrections 
stemming from the QED, weak and strong sector of the
Standard Model (SM)~\cite{Jegerlehner:2009ry,Jegerlehner:2017gek}. 
Hence, the comparison between theory and experiment provides a very stringent 
test of the SM and a deviation from the SM expectation is a monitor for the detection of 
possible New Physics signals.

There is a long-standing and puzzling muon $g-2$ discrepancy between the measured value and the
theoretical prediction, which presently exceeds the $3\sigma$ level. The current status of the SM theoretical prediction for the muon $g-2$
has been very recently reviewed in Ref.~\cite{Aoyama:2020ynm}. 

 Two new experiments,
{\it i.e.} the presently running E989 experiment at Fermilab~\cite{Grange:2015fou,Venanzoni:2014ixa} and the E34 experiment under 
development at J-PARC~\cite{Iinuma:2011zz,Mibe:2011zz}, are expected to improve the current accuracy by a factor of four. 
This calls for a major effort on the theory side in order to reduce the uncertainty in the 
SM prediction, which is dominated by non-perturbative strong interaction effects 
as given by the leading order hadronic correction, $a_\mu^\text{HLO}$, and the
hadronic light-by-light contribution.

Traditionally, $a_\mu^\text{HLO}$ has been computed via
a dispersion integral of the hadron production cross section in electron-positron 
annihilation at low energies~\cite{Jegerlehner:2018zrj,Keshavarzi:2018mgv,Davier:2019can}. 
Lattice QCD calculations are providing
alternative evaluations of the leading order hadronic 
contribution~\cite{DellaMorte:2017dyu,Borsanyi:2017zdw,Wittig:2018r,Meyer:2018til,Blum:2018mom,Giusti:2018mdh,Giusti:2020efo}.
Very recently the BMW collaboration has presented a precise determination
of $a_\mu^\text{HLO}$ with an uncertainty of
$0.7$\%~\cite{Borsanyi:2020mff}, a central value larger than the ones obtained
via dispersive approaches by about $2\sigma$ and in agreement with
the BNL experimental determination~\cite{Bennett:2006fi}.
New and promising algorithms are being developed for lattice QCD
evaluation of $a_\mu^\text{HLO}$ to reduce statistical fluctuations~\cite{DallaBrida:2020cik}.
To clarify this situation,
alternative and independent methods for the evaluation of $a_\mu^\text{HLO}$ are
therefore more than welcome. 

Following ideas first put forward in Refs.~\cite{Lautrup:1969uk,Lautrup:1971jf} for
the evaluation of $a_\mu$ as an integral over the photon
vacuum-polarization function at negative $q^2$,
a novel approach has been recently proposed to derive $a_\mu^\text{HLO}$
from a measurement of the effective electromagnetic coupling constant
in the space-like region via scattering data~\cite{Calame:2015fva}.
Shortly afterwards, the elastic scattering of high-energy muons on atomic electrons 
has been identified as an ideal process for such a
measurement~\cite{Abbiendi:2016xup}~\footnote{A method to measure the running of the
  QED coupling in the space-like region using small-angle Bhabha scattering was proposed
  in Ref.~\cite{Arbuzov:2004wp} and applied to LEP data by the OPAL
  Collaboration~\cite{Abbiendi:2005rx}.} and a new experiment, MUonE,
has been proposed at CERN to measure the differential cross section of 
$\mu e$ scattering as a function of the space-like squared momentum
transfer~\cite{MUonE:LoI}.
In order for this new determination of $a_\mu^\text{HLO}$ to be competitive with
the traditional dispersive approach, the uncertainty in the measurement of the 
$\mu e$ differential cross section must be of the order of 10ppm. This 
experimental target requires high-precision predictions 
for $\mu e$ scattering, including all the relevant radiative corrections.

The extreme accuracy of MUonE demands for theoretical progress in the
calculation of QED corrections to muon-electron scattering, resulting in 
next-generation Monte Carlo (MC) tools. For ultimate data analysis, the latter 
will have to include the full set of NNLO QED corrections combined with 
the leading (and, eventually, next-to-leading) logarithmic contributions due
to multiple photon radiation. Quite recently, a number of steps have been already taken 
to achieve this goal. In Ref.~\cite{Alacevich:2018vez}, the full set of NLO QED and one-loop weak corrections
was computed without any approximation and implemented in a fully exclusive 
MC generator. It is presently being used for simulation studies of MUonE events in the
presence of QED radiation. Important results were also obtained at NNLO accuracy
in QED. The master integrals for the two-loop planar and non-planar four-point Feynman 
diagrams were computed in Refs.~\cite{DiVita:2018nnh,Mastrolia:2017pfy}, 
by setting the electron mass to zero while retaining full dependence on the muon mass. 
A general procedure to extract leading electron mass terms for processes with large masses, 
such as muon-electron scattering, from the corresponding massless
amplitude was given in Ref.~\cite{Engel:2018fsb}
and supplemented with a subtraction scheme for QED calculations with massive fermions
at NNLO accuracy~\cite{Engel:2019nfw}. The two-loop hadronic corrections to $\mu e$ scattering 
were computed in Refs.~\cite{Fael:2019nsf,Fael:2018dmz}. Also possible contamination from
New Physics effects has been studied in Refs.~\cite{Masiero:2020vxk,Dev:2020drf}.
A comprehensive review of the current theoretical knowledge of the
muon-electron scattering cross section for MUonE kinematical conditions
has been published in Ref.~\cite{Banerjee:2020tdt}. 

In this paper, we present the calculation of the fixed-order NNLO QED
photonic corrections and its implementation in a fully fledged MC
generator (\textsc{Mesmer}, Muon Electron Scattering with Multiple
Electromagnetic Radiation~\footnote{The MC code is not public yet, but we
plan to make it available on a public repository soon.}), which represents the first step towards the inclusion
of the full set of NNLO corrections matched to multiple photon emission.
In Section~\ref{sec:nnlo_singleleg}
the gauge invariant subset of photonic corrections
on the single fermionic line (electron and muon) are discussed. These
contributions are exact at NNLO accuracy, including all mass terms. 
The extension of the calculation to include the virtual amplitudes involving
the interference between the
electron and muon line is discussed in Section~\ref{sec:nnlo_full-YFS}.
In this case the two-loop diagrams where at least two virtual 
photons connect the electron and muon lines are not completely known yet.
However, thanks to the universality structure of the infrared (IR)
contributions, the IR part of these NNLO virtual contributions
can be taken into account by means of the classical
Yennie-Frautschi-Suura (YFS) approach~\cite{Yennie:1961ad}.
The real radiation matrix elements and phase space are exact, including all
fermion mass terms. The described theoretical formulation 
is included in a fully exclusive NNLO MC code, needed
for precision simulations at the MuonE experiment. The structure
of the code is completely general and can be easily extended to include
the nowadays missing exact NNLO contributions, thus removing the
present source of approximation.

We performed several cross checks against the
independent calculation of Ref.~\cite{Banerjee:2020rww} for NNLO radiation
stemming from the electron leg, finding perfect agreement. Thanks to this 
common effort, we agreed with the authors
of Ref.~\cite{Banerjee:2020rww} to proceed in parallel and make our results
public at the same time.

By means of the developed MC code we illustrate in
Section~\ref{sec:numerics} numerical results relevant for
typical running condition and event selections of the MUonE experiment. 
In particular, in Section~\ref{sec:numerics-from-lepton-line} 
we show the impact of the exact NNLO photonic corrections due to
electron and muon radiation on the main observables of interest for the
MUonE experiment. 
In Section~\ref{sec:numerics-full} we consider the impact of the
approximate full NNLO photonic corrections on a subset of the  same observables.
In order to roughly estimate the uncertainty of the YFS approximation to
the full NNLO amplitudes, a comparison of YFS at NLO is performed
against the exact NLO calculation of Ref.~\cite{Alacevich:2018vez}.
A summary and future prospects are given in Section~\ref{sec:summary}. 
The work presented in this paper represents the first fundamental step
towards the implementation of a fully-fledged MC generator including the
complete set of NNLO corrections matched to multiple photon emission.

\section{Electron (and muon) line radiation}
\label{sec:nnlo_singleleg}
The complete set of NNLO QED corrections along the electron line entering the $\mu e \to \mu e$ process 
consists of three parts with contributions due to virtual and real photons. The virtual corrections
give rise to ultraviolet divergencies that are computed in Dimensional
Regularization (DR) and
renormalized using the on-shell renormalization scheme. 
\begin{figure}[hbtp]
\begin{center}
\includegraphics[width=0.95\textwidth]{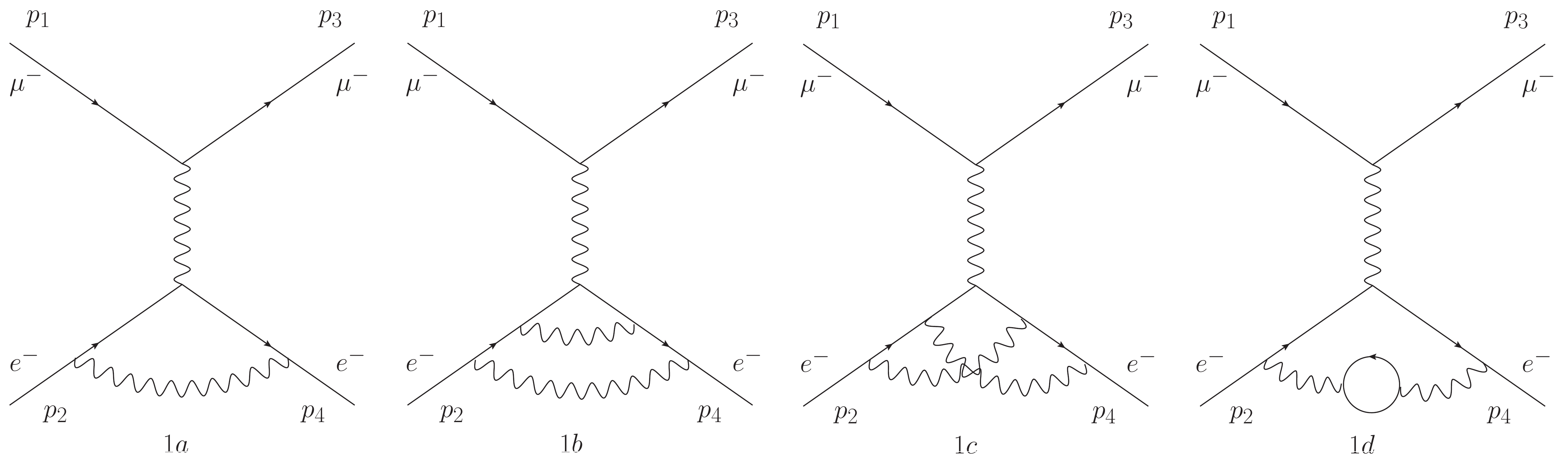}
\end{center}
\caption{Virtual QED corrections to the electron line 
in $\mu e \to \mu e$ scattering. One-loop correction (diagram 1a); 
sample topologies for the two-loop corrections (diagrams 1b-1d). 
The blob in diagram 1d denotes an electron loop insertion. On-shell 
scheme counterterms are understood.}
\label{fig:fig1}
\end{figure}
The three parts contributing to a given differential cross section at NNLO are the following ones:
\begin{itemize}

\item the $d \sigma_\text{virt.}^{\alpha^2}$ {\it two-loop} cross section, that
  consists of squared absolute value of diagrams 1a shown in
  Fig.~\ref{fig:fig1} (summed with relative counterterms)
and irreducible {two-loop contributions} due to diagrams as in 1b-1d in Fig.~\ref{fig:fig1} in interference with the tree-level diagram;

\item the $d \sigma_{1\gamma}^{\alpha^2}$ cross section, that corresponds to the 
{\it one-loop corrections to single photon} emission given by diagrams as 
in 2a-2b in Fig.~\ref{fig:fig2};

\item the $d \sigma_{2\gamma}^{\alpha^2}$ cross section associated to the 
{\it double bremsstrahlung}
process $\mu e \to \mu e + \gamma\gamma$ given by contributions as in 2c-2d in Fig.~\ref{fig:fig2}.
\end{itemize}
\begin{figure}[hbtp]
\begin{center}
\includegraphics[width=0.95\textwidth]{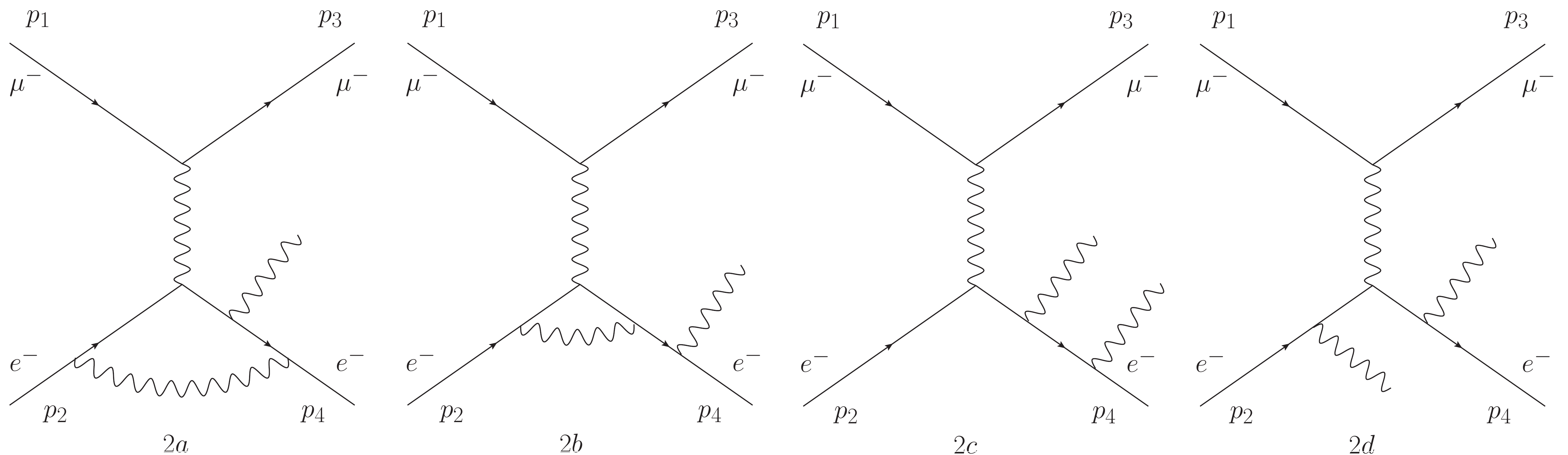}
\end{center}
\caption{Sample diagrams for the one-loop QED corrections to single photon emission (diagrams 2a-2b); 
sample diagrams for the double bremsstrahlung process (diagrams 2c-2d).}
\label{fig:fig2}
\end{figure}

All the above contributions are infrared-divergent quantities and we
choose to regularize IR singularities by assigning a vanishingly small mass $\lambda$ to the photon in the
computation of virtual and real contributions.
Then we introduce a soft-hard slicing separator $\omega_s$, so that
it acts as a fictitious energy resolution parameter of the photon 
radiation phase space, which is split into three sectors: 
the region labelled as $(0\gamma, \text{hard})$ corresponds to the region 
of unresolved radiation up to $\omega_s$,
the domain labelled as $(1\gamma, \text{hard})$ 
corresponds to the region with one resolved photon (with energy $> \omega_s$)
and additional unresolved radiation up to $\omega_s$,
the domain $(2\gamma, \text{hard})$ corresponds to the
region with two resolved photons, where both of them have energy larger than
$\omega_s$. According to the above described splitting, the
pure ${\cal O}(\alpha^2)$ contribution to the cross section
can be rewritten as follows:
\begin{equation}
d\sigma^{\alpha^2} = d \sigma_{0\gamma, \text{hard}} (\omega_s) + 
d \sigma_{1\gamma, \text{hard}} (\omega_s) + d \sigma_{2\gamma, \text{hard}} (\omega_s)\, , 
\label{eq:full}
\end{equation}
where 
\begin{align}
d\sigma_{0\gamma, \text{hard}} (\omega_s) &=
d \sigma_{0\gamma\, \text{s};\,  0\gamma\, \text{h}}^{2\gamma\, \text{virt}}(\lambda) + 
d \sigma_{1\gamma\, \text{s};\, 0\gamma\, \text{h}}^{1\gamma\, \text{virt}} (\lambda,\omega_s) +
d \sigma_{2\gamma\, \text{s}; \, 0\gamma\, \text{h}}^{0\gamma\, \text{virt}} (\lambda,\omega_s) \label{eq:sv} \\ 
d \sigma_{1\gamma, \text{ hard}} (\omega_s) &=
d \sigma_{0\gamma\, \text{s}; \, 1\gamma\, \text{h}}^{1\gamma\, \text{virt}} (\lambda,\omega_s)
+ d \sigma_{1\gamma\, \text{s};\, 1\gamma\, \text{h}}^{0\gamma\, \text{virt}} (\lambda,\omega_s)
\label{eq:hnlo} \\
d \sigma_{2\gamma, \text{ hard}} (\omega_s) &= d \sigma_{0\gamma\, \text{s};\, 2\gamma\, \text{h}}^{0\gamma\, \text{virt}} (\omega_s)
\label{eq:hhnlo}
\end{align}
and the subscripts and superscripts of Eqs.~(\ref{eq:sv})-(\ref{eq:hhnlo})
indicate explicitly the number of virtual
$(\text{virt})$, soft $(\text{s})$ and hard $(\text{h})$ photons. 
In Eq. (\ref{eq:full}), the cross section
$d \sigma_{1\gamma, \text{hard}} (\omega_s)$  
stands for the one-loop correction to the single bremsstrahlung 
with emission of a photon of energy greater 
than $\omega_s$. Analogously,
$d \sigma_{2\gamma, \text{hard}} (\omega_s)$ is 
the cross section corresponding to the radiation of two hard photons.
The dependence on the photon mass parameter has been made explicit in
Eqs.~(\ref{eq:sv}) and (\ref{eq:hnlo}). Each term entering
Eq.~(\ref{eq:full}) is independent of $\lambda$.
The radiation of real soft photons with energy smaller than $\omega_s$ 
is included by means of analytical eikonal
factors~\cite{tHooft:1978jhc,Denner:1991kt,Denner:2019vbn}. 
The sum of the right-hand side of Eq.~(\ref{eq:full})
is also independent of $\omega_s$, 
so that Eq.~(\ref{eq:full}) can be used to take into account any realistic
experimental event selection.

The master formula with up to NNLO accuracy implemented in our MC generator
reads as follows
\begin{equation}
  d\sigma^{\text{NNLO}} = d\sigma^{\alpha^0} + d\sigma^{\alpha^1}
    + d\sigma^{\alpha^2}\, ,
\label{eq:masterMC}
\end{equation}
where $d\sigma^{\alpha^0}$ and $d\sigma^{\alpha^1}$ represent
the tree-level and the NLO contributions, respectively. For
definiteness and later use, we also define
\begin{align}
  d\sigma^{\text{LO}} & = d\sigma^{\alpha^0} \, , \label{eq:masterMClo}\\
  d\sigma^{\text{NLO}} & = d\sigma^{\alpha^0} + d\sigma^{\alpha^1}\, . \label{eq:masterMCnlo}
\end{align}

For the calculation of the radiative corrections with one virtual photon
and the double bremsstrahlung process, all the Feynman diagrams were manipulated with the help of the 
algebraic manipulation program \textsc{Form}~\cite{Kuipers:2012rf,Ruijl:2017dtg}, 
keeping full dependence on the lepton masses.
The evaluation of one-loop tensor coefficients and scalar 
functions is performed using the package
\textsc{Collier}~\cite{Denner:2016kdg}~\footnote{Several cross-checks have been performed with
  the package~\textsc{LoopTools}~\cite{Hahn:2010zi,Hahn:1998yk}
  finding very good agreement.}. 
As a cross-check, we compared our QED calculation of the one-loop Dirac and Pauli form factors with the Abelian limit of 
the QCD results given in Ref.~\cite{Bernreuther:2004ih} (see also \cite{Gluza:2009yy}), converting the soft-IR pole in DR to
the logarithm involving the photon mass $\lambda$ according to the
rule $\left(4 \pi \mu^2\right)^\epsilon\Gamma(1+\epsilon)/\epsilon\to\ln\lambda^2$~\cite{Denner:2019vbn}, finding perfect agreement. 
For the virtual two-loop corrections, 
we resorted to the calculation of the ${\cal O} (\alpha^2)$ QED form factors 
of Ref.~\cite{Mastrolia:2003yz}, where IR divergences are parameterized
in terms of the fictitious photon mass $\lambda$ and all finite lepton 
mass effects are kept~\footnote{We thank P. Mastrolia and A. Primo for
  providing us with updated exact expressions of the form factors.}.
In the latter, the result is expressed in terms of Harmonic Polylogarithms
and we checked that the asymptotic expansion for large momentum transfer
agrees with the high-energy limit of the two-loop form factors
given in Refs.~\cite{Berends:1987ab,Burgers:1985qg}, up to constant
terms. It is worth mentioning that the QCD NNLO form factors have been calculated
also in Ref.~\cite{Ablinger:2017hst}, with partial results relevant
for the full three-loop form factor calculation in Refs.~\cite{Blumlein:2019oas,Ablinger:2018zwz,Blumlein:2018tmz,Ablinger:2018yae}.

The ${\cal O} (\alpha^2)$ QED form factors also contain the contribution of the
diagram with a vacuum polarization insertion on the vertex photonic
correction (see Figure~\ref{fig:fig1}, diagram 1d), due to the same lepton 
of the external legs. This contribution is enhanced
by a term proportional to $\alpha^2 L^3$, where
$L= \ln \left(\frac{-t}{m^2}\right)$ represents the collinear logarithm,
which is, in any case, exactly cancelled by the emission of a real leptonic pair, when
the latter is integrated on the available phase
space~\cite{Montagna:1998vb}. For this reason
we define our NNLO photonic corrections by removing the contribution
due to the fermionic loop, which will be investigated in the future with
real leptonic pair emission. This $N_f=1$ contribution to the Dirac and
Pauli form factors has been computed independently in DR  and found to agree
with Eqs.~(94) and~(95) of Ref.~\cite{Bonciani:2004gi}. The final result
is finite and independent of the regularization scheme adopted
in the intermediate steps. 

\section{YFS-inspired approximation of the full two-loop amplitude}
\label{sec:nnlo_full-YFS}
In this section, we discuss how the full two-loop virtual
QED corrections can be approximated to catch its complete IR
structure. The aim is to implement a full, although approximate,
two-loop QED correction to the process $\mu e\to\mu e$ which includes correctly
at least all the IR parts. In particular, we are approximating only the
diagrams obtained by inserting a virtual photon into a QED one-loop box
diagram, all the rest of the corrections being exact. We remark that
also the full one-loop corrections to the radiative processes $\mu^\pm
e^-\to\mu^\pm e^-\gamma$ are needed and calculated exactly, including box and pentagon
diagrams. The latter are calculated with the techniques described
above for one-loop corrections~\footnote{For the full one-loop QED corrections 
to the process of single-photon emission, we find agreement with the
matrix elements calculated  by using the \textsc{Recola-Collier} package~\cite{Actis:2016mpe,Denner:2016kdg}.
Particular attention is being
  paid in order to make the pentagon amplitudes numerically stable in
  phase-space regions where the
  external photon momentum tends to be soft. The details are beyond the
  scope of the present paper and will be possibly discussed elsewhere.}.

The full two-loop virtual amplitude is built-up exploiting the
YFS analysis of IR divergencies in
QED~\cite{Yennie:1961ad}. The starting point are Eqs.~(2.2a-c) of the
original YFS paper, where the exact $n$-loop amplitude is formally written
extracting order by order IR factors in an iterative way. Stopping at
two-loop, their equations can be written as

\begin{align}
\mathcal{M}^{\alpha^0}&=\mathcal{T} \nonumber \\
\mathcal{M}^{\alpha^1}&=Y\mathcal{T} + \mathcal{M}^{\alpha^1,\text{R}} \nonumber\\
\mathcal{M}^{\alpha^2}&=\frac{1}{2} Y^2\mathcal{T} + Y\mathcal{M}^{\alpha^1,\text{R}} + \mathcal{M}^{\alpha^2,\text{R}} \nonumber\\
                   &=-\frac{1}{2} Y^2\mathcal{T} + Y\mathcal{M}^{\alpha^1} + \mathcal{M}^{\alpha^2,\text{R}}
\label{eq:YFS2loop}
\end{align}
where $\mathcal{T}$ represents the tree level amplitude,
$\mathcal{M}^{\alpha^n}$ are the exact $n$-loop amplitudes, $Y$ (of order $\alpha$) is the YFS IR
virtual factor and the superscript R indicates the non-IR remnant of
the amplitude. We stress that IR divergencies are confined only in the
terms containing at least one $Y$ factor.

Restricting to the process under consideration and using momenta
labelling as in Fig.~\ref{fig:fig1}, the IR factor $Y$ explicitly reads
(see also Refs.~\cite{Jadach:2000ir,Schonherr:2008av,Krauss:2018djz,Linten:2018dpx})
\begin{equation}
  Y = \sum_{i,j=1,4}^{j\ge i}Y_{ij} = Y_e + Y_\mu + Y_{e\mu}
  \label{eq:YIR}
\end{equation}
where
\begin{equation}
  Y_{ij} = \begin{cases}
          \frac{1}{8}\frac{\alpha}{\pi}Q_i^2\left[B_0\left(0,m_i^2,m_i^2\right)-
            4m_i^2C_0\left(m_i^2,0,m_i^2,\lambda^2,m_i^2,m_i^2\right)\right]
          & \text{for }i=j \\
          \frac{\alpha}{\pi}Q_iQ_j\vartheta_i\vartheta_j\left[p_i\cdot
            p_j\;C_0\left(m_i^2,(\vartheta_ip_i+\vartheta_ip_j)^2,m_j^2,\lambda^2,m_i^2,m_j^2\right)+\right. &  \\
            \qquad\qquad+\left.\frac{1}{4}B_0\left((\vartheta_ip_i+\vartheta_jp_j)^2,m_i^2,m_j^2\right)
            \right] & \text{for }i\neq j
  \end{cases}
\label{eq:YFSfactors}
\end{equation}
\begin{align}
Y_e &= Y_{24} + Y_{22} + Y_{44}\nonumber\\
Y_\mu &= Y_{13} + Y_{11} +Y_{33}\nonumber\\
Y_{e\mu} &= Y_{12} + Y_{14} + Y_{23} + Y_{34}
\label{eq:YeYmuYemu}
\end{align}

In Eq.~(\ref{eq:YFSfactors}), $Q_i$ is the charge of particle $i$ in
positron charge units, $\vartheta_i=-1\; (1)$ for an incoming (outgoing) fermion
and the arguments of the scalar one-loop functions $B_0$ and $C_0$ follow
\textsc{Collier}/\textsc{LoopTools} conventions. We stress that $Y_a$ with
$a=e,\mu,e\mu$ are the factors which factorize the IR divergence
of the diagrams obtained by dressing the underlying amplitude
with an extra virtual photon attached to the electron
($a=e$) or muon ($a=\mu$) line or connecting them ($a=e\mu$). An analogous
separation holds for the exact one-loop amplitude $\mathcal{M}^{\alpha^1}$ and
its non-IR remnant $\mathcal{M}^{\alpha^1,\text{R}}$  of Eq.~(\ref{eq:YFS2loop}), {\it i.e.} we can write
$\mathcal{M}^{\alpha^1\text{(,R)}} = \mathcal{M}^{\alpha^1\text{(,R)}}_e + \mathcal{M}^{\alpha^1\text{(,R)}}_\mu + \mathcal{M}^{\alpha^1\text{(,R)}}_{e\mu}$.

We can now approximate the complete two-loop virtual amplitude
$\mathcal{M}^{\alpha^2}$ by writing
\begin{align}
  {\mathcal{M}^{\alpha^2}} \simeq \widetilde{\mathcal{M}}^{\alpha^2} &=
  \mathcal{M}^{\alpha^2}_e + \mathcal{M}^{\alpha^2}_\mu + \mathcal{M}^{\alpha^2}_{e\mu,\,{\text{1L}\times\text{1L}}}\nonumber\\
&+ \frac{1}{2}Y^2\mathcal{T} + Y\mathcal{M}^{\alpha^1,\text{R}} \nonumber\\
&- \left(\frac{1}{2}Y_e^2\mathcal{T} + Y_e\mathcal{M}^{\alpha^1,\text{R}}_e \right)
 - \left(\frac{1}{2}Y_\mu^2\mathcal{T} + Y_\mu\mathcal{M}^{\alpha^1,\text{R}}_\mu \right)\nonumber\\
&  - \left(Y_eY_\mu\mathcal{T} + Y_e\mathcal{M}^{\alpha^1,\text{R}}_\mu + Y_\mu\mathcal{M}^{\alpha^1,\text{R}}_e\right), 
\label{eq:YFSapproxfull}
\end{align}
where in the second line of Eq.~(\ref{eq:YFSapproxfull}) we have put
$\mathcal{M}^{\alpha^2,\text{R}}=0$. 
With little algebra, the previous definition can be also recast in the following form
\begin{align}
  \widetilde{\mathcal{M}}^{\alpha^2} &=
  \mathcal{M}^{\alpha^2}_{e} + \mathcal{M}^{\alpha^2}_\mu + \mathcal{M}^{\alpha^2}_{e\mu,\,{\text{1L}\times\text{1L}}}\nonumber\\
& + \frac{1}{2}Y_{e\mu}^2\mathcal{T} + Y_{e\mu}\left(Y_e+Y_\mu\right)\mathcal{T} + (Y_e+Y_\mu)\mathcal{M}^{\alpha^1,\text{R}}_{e\mu} + Y_{e\mu} M^{\alpha^1,\text{R}}.
\label{eq:YFSapproxfull2}
\end{align}
In Eq.~(\ref{eq:YFSapproxfull}), in the first row,
$\mathcal{M}^{\alpha^2}_e$ ($\mathcal{M}^{\alpha^2}_\mu$) represents the set of two-loop diagrams
where the two virtual photons are both attached to the
electron (muon) line, whereas
$\mathcal{M}^{\alpha^2}_{e\mu,\,{\text{1L}\times\text{1L}}}$ represents the
diagrams where one
photon is on the electron line and the other one on the muon line~\footnote{This
  set can be easily computed exactly, for instance by using one-loop renormalized vertex
  form factors.}. The second row contains the full IR part of
Eq.~(\ref{eq:YFS2loop}). The third and fourth rows finally subtract from the second
one the IR content already accounted for in
$\mathcal{M}^{\alpha^2}_e + \mathcal{M}^{\alpha^2}_\mu + \mathcal{M}^{\alpha^2}_{e\mu,\,{\text{1L}\times\text{1L}}}$. In other words, the
approximation~$\widetilde{\mathcal{M}}^{\alpha^2}$ misses only the non-IR remnant of
the two-loop diagrams where at least two photons connect the
electron and muon lines, which are not known yet for the process
under consideration. This is made possibly more explicit in the
equivalent form
of Eq.~(\ref{eq:YFSapproxfull2}), where each term in the second row, approximately accounting for
the missing contributions, has a factor with an ``$e\mu$'' subscript. Despite being incomplete, the exact IR structure of the
full two-loop amplitude, with massive electron and muon, is present in
Eqs.~(\ref{eq:YFSapproxfull}) and~(\ref{eq:YFSapproxfull2}): on the one side 
this can be used to cross-check the IR part of the final calculation once
it is available, on the other side it can be implemented into the Monte
Carlo generator for phenomenological studies. We stress that,
when we add $2\Re(\widetilde{\mathcal{M}}^{\alpha^2}\mathcal{T}^*)$
to the rest of the exact $\mathcal{O}(\alpha^2)$ contributions,
the sum is perfectly independent of the fictitious
photon mass $\lambda$ and thus the result is meaningful.

On the accuracy side, heuristic considerations, also based upon an analogous numerical
analysis at one-loop order, bring us to estimate the size of the
dominant missing non-IR ${\cal O}(\alpha^2)$ corrections to be in the range of a few
$\left(\frac{\alpha}{\pi}\right)^2\ln^2\left(m_\mu^2/m_e^2\right)\simeq
6\times 10^{-4}$, as discussed later.

\section{Numerical results}
\label{sec:numerics}
In this Section, we show and discuss the phenomenological results obtained by
using a fully differential MC code, \textsc{Mesmer}, which implements the theoretical
approach described in Section~\ref{sec:nnlo_singleleg} and Section~\ref{sec:nnlo_full-YFS}. We adopt the parameters and typical
running condition of the MUonE experiment already described in detail
in Ref.~\cite{Alacevich:2018vez}, which we briefly remind here:
the reference frame is the laboratory frame, where the energy
of the incoming $\mu^\pm$ is $E_\mu^\text{beam} = 150$~GeV, the electron
is assumed to be at rest inside a bulk target and thus $\sqrt{s}\simeq0.405541\text{ GeV}$. 
In this kinematical condition,
the collinear logarithms $L_e = \ln (s/m_e^2)$ and 
$L_\mu = \ln (s/m_\mu^2)$
are of the order of $L_e \simeq 13.4$ and $L_\mu \simeq 2.7$, respectively.
We study the numerical impact of QED NNLO photonic corrections to observables
assuming the following event selections
(Setup~2 and Setup~4 of Ref.~\cite{Alacevich:2018vez}): 
\begin{enumerate}
\item  $\vartheta_e,\vartheta_\mu<100\text{\ mrad}$ and $E_e > 1$~GeV
  {(i.e. $t_{ee}\lesssim -1.02\cdot 10^{-3}\text{ GeV}^2$)}. The angular cuts model the typical acceptance 
conditions of the experiment and the electron energy threshold is imposed to
guarantee the presence of two charged tracks in the detector (Setup~1);
\item the same criteria as above, with the additional acoplanarity cut 
  $|\pi - |\phi_e-\phi_\mu||\le 3.5\text{\ mrad}$. We remind the reader
  that this event selection is considered in order to mimic an experimental
  cut which allows to stay close to the elasticity curve given by the
  tree-level relation between the electron and muon scattering angles
  (Setup~2)\footnote{The acoplanarity cut partially removes radiative
  events and thus enhances the fraction of quasi-elastic events, as
  illustrated by Fig. 3 of Ref.~\cite{Alacevich:2018vez}. Here it is
  used merely as a possible cut for selecting elastic events, without
  pretending it to be realistic. Indeed, from the experimental point
  of view, the extraction of the hadronic contribution to the running
  of $\alpha(t)$ could benefit from selecting events in the
  elastic region.},
\end{enumerate}
where $t_{ee} = (p_2 - p_4)^2$, $(\vartheta_e, \phi_e, E_e)$ and $(\vartheta_\mu, \phi_\mu, E_\mu)$
are the scattering and azimuthal angles
and the energy, in the laboratory frame, of the outgoing electron and muon,
respectively. 

For Setup~1 and Setup~3 of Ref.~\cite{Alacevich:2018vez}, with
$E_\gamma > 0.2$~GeV, the NNLO QED photonic effects are similar
in size to the ones presented here and therefore are not considered
for the sake of brevity.

The input parameters for the simulations are set to
\begin{equation}
  \alpha = 1/137.03599907430637 \qquad 
  m_e = 0.510998928\text{ MeV} \qquad m_\mu = 105.6583715\text{ MeV}\nonumber
  \label{eq:inputqedparams}
\end{equation}
and the considered differential observables are the following ones:
\begin{equation}
  \frac{d\sigma}{d t_{ee}}\,, \, \, \, \, \, \, 
  \frac{d\sigma}{d t_{\mu \mu}}\,, \, \, \, \, \, \,
  \frac{d\sigma}{d\vartheta_e}\,, \, \, \, \, \, \,
  \frac{d\sigma}{d\vartheta_\mu}\,, \, \, \, \, \, \,
  \frac{d\sigma}{d E_e}\,, \, \, \, \, \, \,
  \frac{d\sigma}{d E_\mu} \,,
  \label{eq:observables}
  \end{equation}
where $t_{\mu \mu} = (p_1 - p_3)^2$.

For definiteness, we remark that all calculations and simulations are
performed in the center of mass (c.m.) frame and then the momenta are
boosted to the laboratory frame, where the initial-state electron is at rest. 
The soft-hard separator $\omega_s$, not Lorentz invariant, is defined
in the c.m. reference frame and its default value is taken to be
$\omega_s = 10^{-5}\times\sqrt{s}/2$. We checked both at integrated
and differential level that the cross-sections are independent of the
unphysical parameters $\lambda$ and $\omega_s$.

Within the full set of NLO and NNLO QED corrections, there are three gauge-invariant subsets of 
photonic contributions corresponding to:
\begin{enumerate}
\item virtual and real corrections along the electron line;
\item virtual and real corrections along the muon line;
\item the rest of the virtual and real corrections, including box-like
  contributions, up-down interference of real photon radiation and so on.
\end{enumerate}

In Section~\ref{sec:numerics-from-lepton-line} we present
the exact QED NNLO photonic corrections for the distributions of
Eq.~(\ref{eq:observables}) for the electron and muon line separately.
In Section~\ref{sec:numerics-full} we present the results for
our approximation to the full QED NNLO virtual photonic corrections described
in Sect.~\ref{sec:nnlo_full-YFS}~\footnote{We remind the reader that in our approximation any real radiation contribution
is exact, including all finite mass effects.}.
All the figures show  the difference in per cent between the NNLO and NLO predictions
relative to the tree-level differential cross section, {\it i.e.}
\begin{equation}
  \Delta_\text{NNLO}^i \, = \,
  100 \times \frac{d\sigma_\text{NNLO}^i - d\sigma_\text{NLO}^i}{d\sigma_\text{LO}}\, , 
  \label{eq:delta-def}
\end{equation}
where $i = e, \mu, f$ stand for electron-line only, muon-line only
and the full set of NNLO corrections.
For the sake of clarity, all the figures contain insets which display the
NLO corrections w.r.t. the LO predictions, according to 
\begin{equation}
  \Delta_\text{NLO}^i \, = \,
  100 \times \frac{d\sigma_\text{NLO}^i - d\sigma_\text{LO}}{d\sigma_\text{LO}}\, .
  \label{eq:delta_NLO-def}
\end{equation}
In order to quantify the purely photonic effects, 
we do not include any vacuum polarization correction neither in
the NLO nor in the NNLO cross sections.
The absolute values and shapes of the observables can be obtained from
Ref.~\cite{Alacevich:2018vez}. In this Section, figures with two
panels show corrections in Setup~1 on the left and 
in Setup~2 on the right, unless stated otherwise.
\begin{table}[t]
\centering
\begin{tabular}{| c || c | c || c | c |}
\hline
 $\sigma$ ($\mu$b)  & \multicolumn{2}{|c||}{Setup~1} & \multicolumn{2}{|c|}{Setup~2}\\
\hline
& $\mu^{ -}e^{ -}  \to \mu^{ -} e^{ -}$ &$\mu^{ +}e^{ -}  \to \mu^{ +}  e^{ -}$& $\mu^{ -}e^{ -}  \to \mu^{ -} e^{ -}$ &$\mu^{ +}e^{ -}  \to \mu^{ +}  e^{ -}$\\
\hline
\hline
$\sigma_\text{LO}$&\multicolumn{4}{| c |}{$245.038910(1)$}\\
\hline
\hline
$\sigma_\text{NLO}^e$ & \multicolumn{2}{| c ||}{$255.5500(7)$} & \multicolumn{2}{| c |}{$223.4387(6)$} \\
\hline
$\sigma_\text{NLO}^\mu$ & \multicolumn{2}{| c ||}{$244.9707(1)$} & \multicolumn{2}{| c |}{$244.4136(1)$} \\
\hline
$\sigma_\text{NLO}^f$ & 255.1176(5) & 255.8437(5) & 222.8545(3) & 222.7714(3)\\
\hline
\hline
$\sigma_\text{NNLO}^e$ & \multicolumn{2}{| c ||}{$255.5725(5)$} & \multicolumn{2}{| c |}{$224.4796(4)$} \\
\hline
$\sigma_\text{NNLO}^\mu$ & \multicolumn{2}{| c ||}{$244.9706(1)$} & \multicolumn{2}{| c |}{$244.4154(1)$} \\
\hline
$\sigma_\text{NNLO}^f$ & $\mathit{255.205(1)}$ & $\mathit{256.092(1)}$ & $\mathit{224.041(1)}$& $\mathit{224.088(1)}$\\
\hline
\end{tabular}
\caption{\label{tab:tab1}  Cross sections (in $\mu$b) and relative corrections for the 
processes $\mu^- e^- \to \mu^- e^-$ and $\mu^+ e^- \to \mu^+ e^-$, in
the two different setups described in the
text. The symbols 
$\sigma_{(\rm N)(\rm N)\rm LO}^{e/\mu/f}$ stand for 
the cross sections with corrections along the electron line only,
along the muon line only and the full approximate contributions, respectively,
with the perturbative accuracy given by the subscripts. 
The digits in parenthesis correspond to $1\sigma$ MC
error. Italicized numbers in the last row indicate that in this
cross-section the full two-loop amplitude is approximated as
described in Section~\ref{sec:nnlo_full-YFS}.}
\end{table}

Before showing the effects on the distributions, we
quote in Tab.~\ref{tab:tab1}  the integrated cross-sections for the two Setups, showing the
different classes of corrections. The symbols 
$\sigma_{(\rm N)(\rm N)\rm LO}^{e/\mu/f}$ stand for 
the cross sections with corrections along the electron line only,
along the muon line only and the full approximate contributions, respectively,
with the perturbative accuracy given by the subscripts. 
The subsets of corrections coming
from electron or muon line separately are the same for
both processes $\mu^- e^- \to \mu^- e^-$ and $\mu^+ e^- \to \mu^+e^-$,
as expected. For Setup~1 the correction to the cross section along the
electron line amounts to less than 0.01\% with positive sign, with respect
to the NLO prediction, while the correction along the muon line is
below the $10^{-5}$ level, still compatible with zero within the
statistical integration uncertainty. For Setup~2 the correction
along the electron line becomes of the order of $+0.45$\% and the one
along the muon line below the $0.01$\% level. 
The numbers in the last row of the table are italicized
to indicate that these cross-sections are approximate according to the
discussion of Sect.~\ref{sec:nnlo_full-YFS}. Numerically, the full corrections
in Setup~1 amount to about $+0.04$\%, with respect to the NLO prediction, for
the process with $\mu^-$ beam, while it amounts to about $+0.1$\% with
$\mu^+$ in the initial state. Given the inherent uncertainty,
roughly estimated to be of the order
of $0.06$\% in Section~\ref{sec:nnlo_full-YFS}, these estimates can not
be considered conclusive yet. When including the acoplanarity cut (Setup~2), 
the approximate NNLO corrections become of the order of $0.5$\%, because of
IR effects enhancement, thus being more robust in view of the
intrinsic uncertainty of the approximate prediction discussed before.

\subsection{Exact NNLO corrections from one leptonic line}
\label{sec:numerics-from-lepton-line}
In this Section we present the results for the relative contribution of the NNLO photonic corrections
to the differential distributions according to Eq.~(\ref{eq:delta-def}). In particular,
we consider the exact NNLO corrections due to electron and muon radiation, which represent different gauge invariant subsets
of the whole NNLO corrections.
The correction along the electron line is particularly important,
because, as shown in Ref.~\cite{Alacevich:2018vez}, already at NLO level it
is enhanced w.r.t. the radiation from the muon line, as a consequence
of the small ratio $m_e/m_\mu$. Moreover, it gives the bulk of the correction
for some observables or in particular phase-space regions. 

\begin{figure}[t]
\begin{center}
\includegraphics[width=0.95\textwidth]{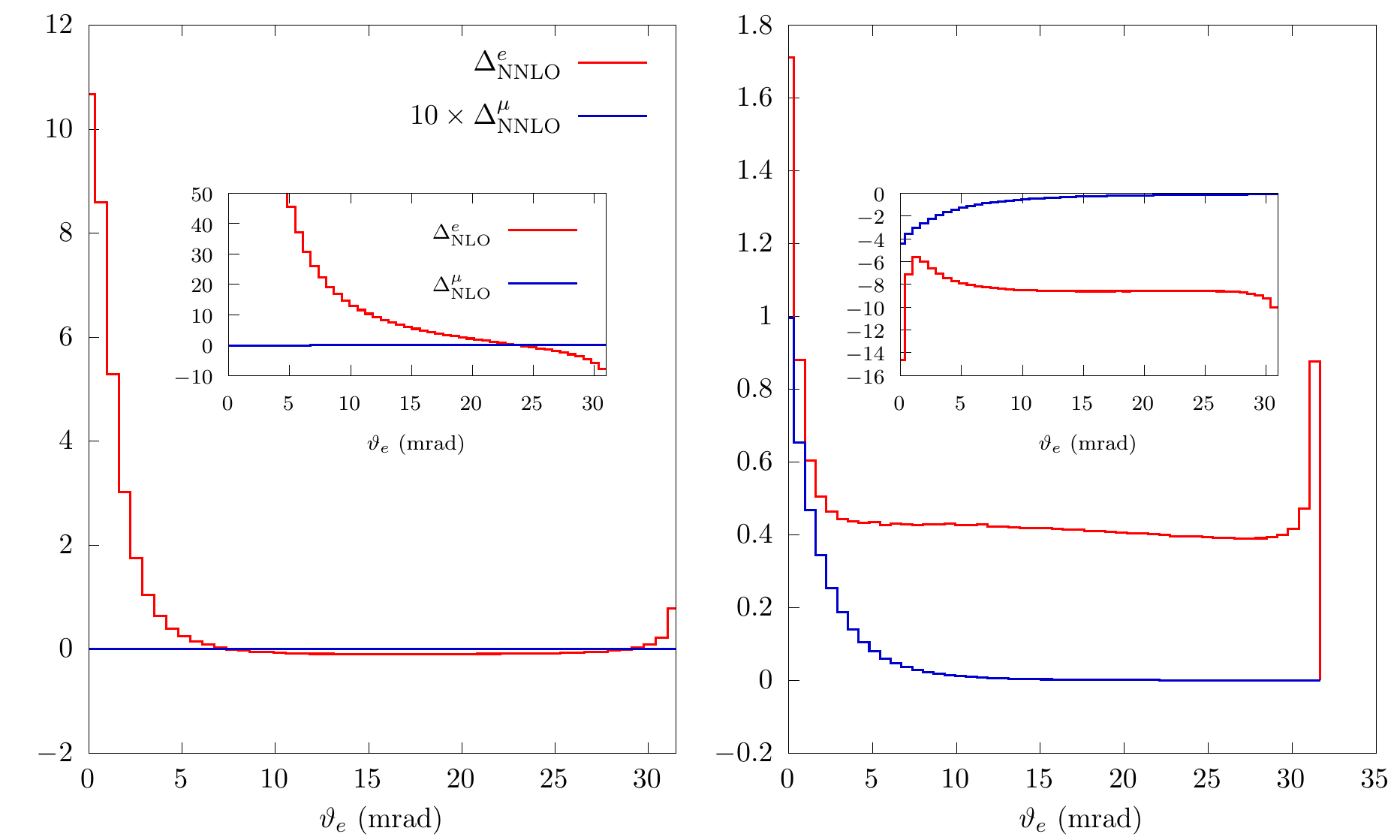}
\end{center}
\caption{\label{Fig:ethlab-0-1} Red histograms: NNLO corrections, 
  according to Eq.~(\ref{eq:delta-def}), along the electron line,
  for the $\mu^\pm e^- \to \mu^\pm e^- $ processes, as a function of
  the electron scattering angle in the laboratory frame, $\vartheta_e$;
  blue histogram: $10\times$NNLO corrections along the muon line. 
  Left panel: differential distribution in the presence of acceptance cuts
  only (Setup~1); right panel: the same as in the left panel with
  the additional acoplanarity cut of Setup~2.
  The insets report the NLO corrections,
  according to Eq.~(\ref{eq:delta_NLO-def}); the blue line represents
  directly Eq.~(\ref{eq:delta_NLO-def}) without any multiplicative factor.
}
\end{figure}
Figure~\ref{Fig:ethlab-0-1} shows the effects of NNLO corrections, according to Eq.~(\ref{eq:delta-def}),
for the observable $d\sigma / d\vartheta_e$. 
In Setup~1, the correction along the electron line
(red histogram) is flat over a large portion of scattering angles larger than
about $5$~mrad, being of the order of a few $0.1\%$ at most,
and displays a steep increase for $\vartheta_e \to 0$, where it 
reaches the order of $10\%$.
As can be seen from the inset, the electron scattering angle distribution
$d\sigma / d\vartheta_e$ is quite
sensitive to NLO corrections, in particular for small
values of $\vartheta_e$, where the NLO correction becomes particularly enhanced
as due to hard bremsstrahlung effects and because 
the tree-level differential cross section tends to zero. As a consequence, 
the increase of the NNLO correction for $\vartheta_e\to 0$ is not surprising.
Contrarily to the correction along the electron line, the one along the
muon line (blue histogram, multiplied by a factor of 10)
is flat over the whole $\vartheta_e$ angular spectrum and invisible on
the plot scale.
The shape of the corrections remains qualitatively the same also in the
presence of the additional acoplanarity cut of Setup~2, even
if the size of the correction along the electron line
becomes larger ($\sim +0.4\%$) for a wide range of $\vartheta_e$.
This is a consequence of the
effects of the IR enhancement introduced effectively by the acoplanarity cut.
Indeed the NLO correction grows to negative values of the order of
$-8\%$. In Setup~2, the NNLO correction due to electron radiation is about $+1.7\%$ for $\vartheta_e \to 0$~mrad
and amounts to $0.8\%$ at the opposite kinematical boundary.
Even in the presence of the acoplanarity cut, the NNLO correction along the
muon line remains flat over the entire spectrum, increasing to about $+0.1\%$ for $\vartheta_e \to 0$, consistently
with the behaviour of the NLO correction shown in the inset.

\begin{figure}[t]
\begin{center}
\includegraphics[width=0.95\textwidth]{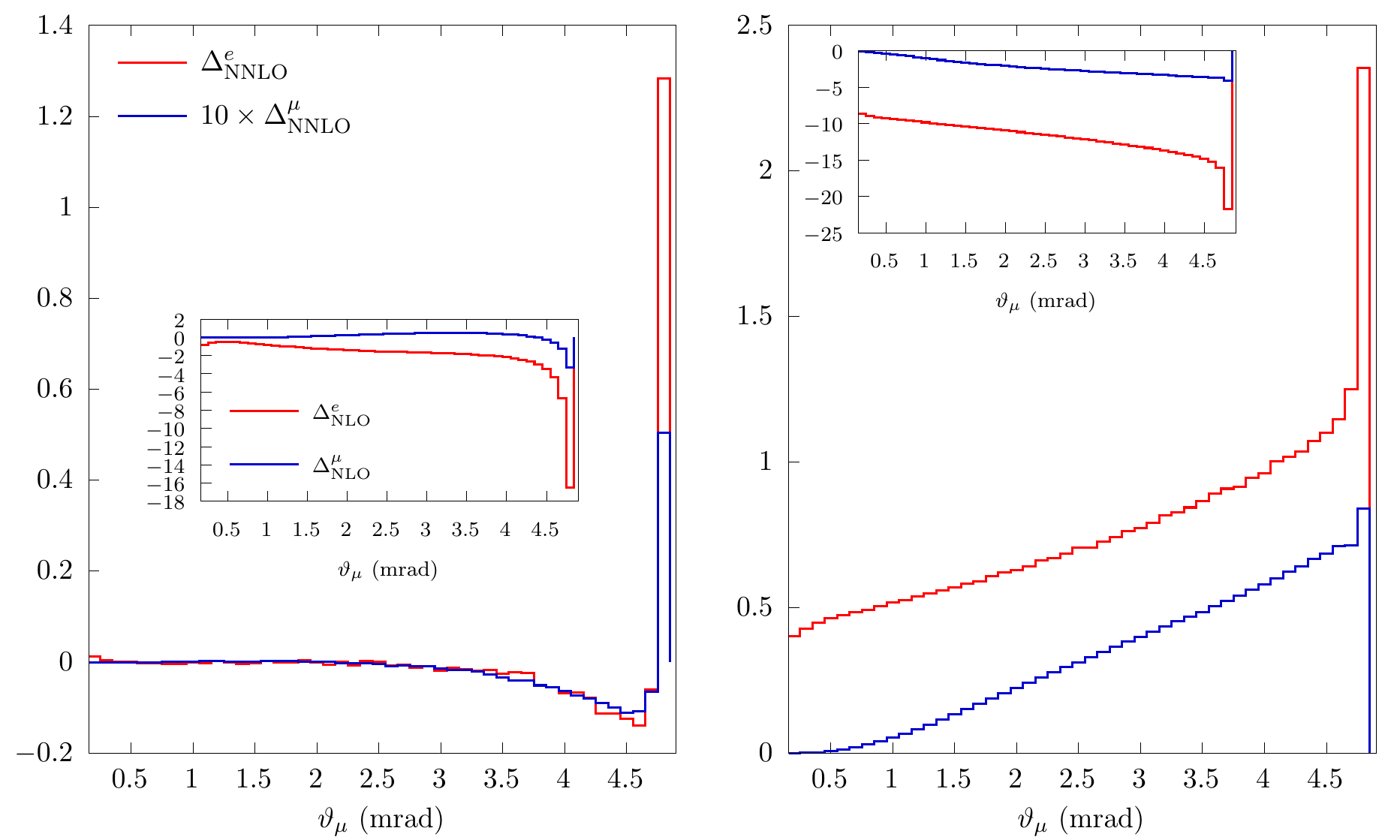}
\end{center}
\caption{\label{Fig:mthlab-0-1} The same as Fig.~\ref{Fig:ethlab-0-1}
  for the muon scattering angle $\vartheta_{\mu}$.}
\end{figure}
Figure~\ref{Fig:mthlab-0-1} shows the effects of NNLO corrections
for the observable $d\sigma / d\vartheta_\mu$~\footnote{In this case, we do not
show muon scattering angles $\vartheta_\mu<0.16\text{ mrad}$ because the
corrections are huge and would spoil the readability of the plots. The reason
for this feature is that the LO prediction becomes negligible in the above
angular range, as a consequence of the applied acceptance cuts, while
the single- and double-radiative events populate the region.}.
In Setup~1 (left panel), the NNLO correction along the electron line 
(red histogram) has a non-trivial shape, remaining of the order
of $0.01\%$  and dropping to about $-0.14\%$ from $2.5$~mrad to
$4.5$~mrad, with a sharp increase in the last two bins, reaching
the level of $+1.3\%$ at the kinematical limit.
Interestingly, the NNLO correction along
the muon line (blue histogram) displays the same shape of the
correction along the electron line but with a reduction in size
by a factor of ten (a factor of 30 in the last bin). 
The shapes of the corrections change when
the acoplanarity cut is added (right panel). In this case the correction
along the electron line (red histogram) becomes an increasing function
of $\vartheta_\mu$, ranging from $+0.4\%$ for $\vartheta_\mu$ in the left
corner of the plot
to about $2.4\%$ at the upper kinematical limit. This large and positive
correction can be understood by looking at the inset, which shows that the
NLO correction varies
monotonically between $-8\%$ and $-22\%$, as
enhanced by IR effects introduced by the acoplanarity cut.
The NNLO correction along the muon line displays
qualitatively a similar shape and reaches about $+0.08\%$ at the upper kinematical limit.

\begin{figure}[t]
\begin{center}
\includegraphics[width=0.95\textwidth]{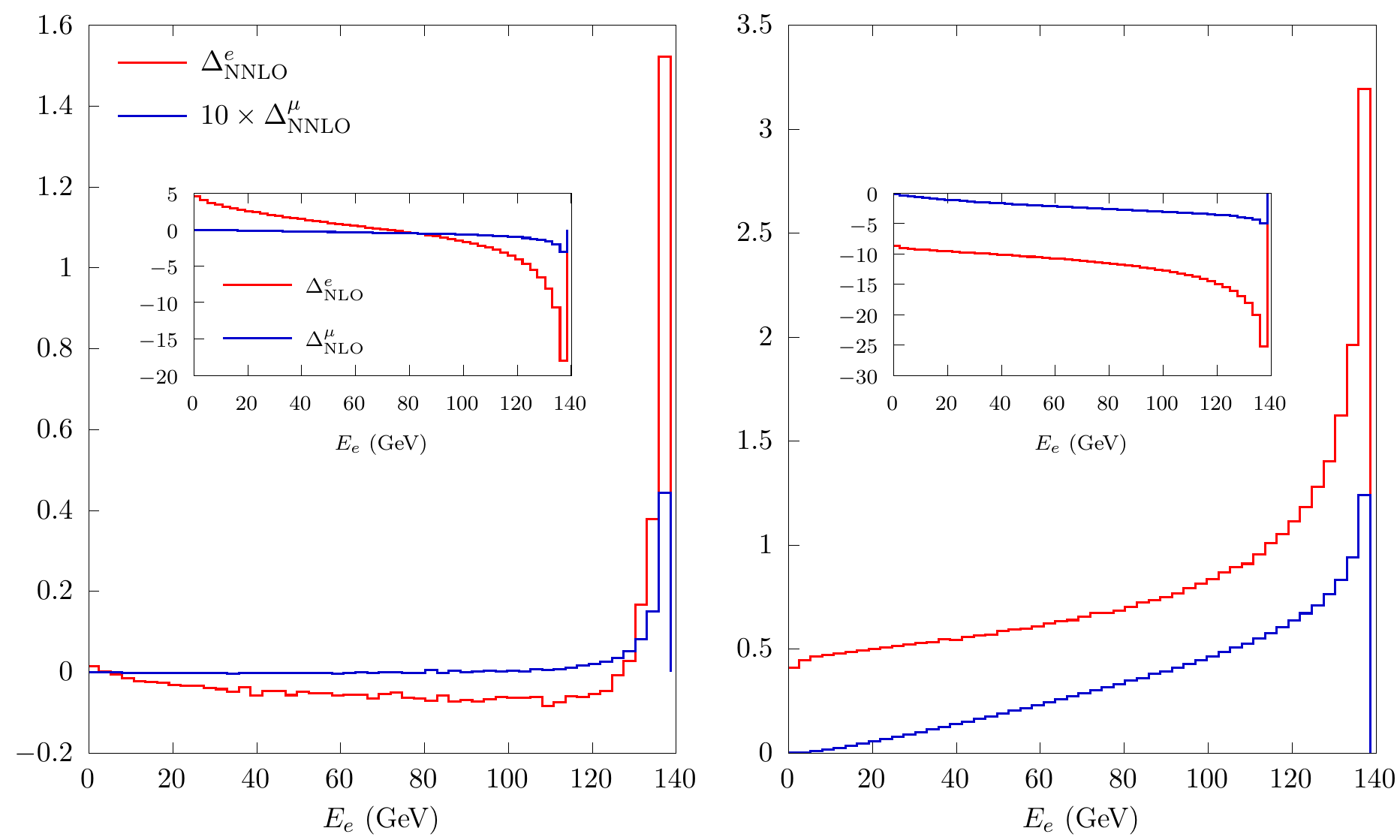}
\end{center}
\caption{\label{Fig:eenlab-0-1} The same as Fig.~\ref{Fig:ethlab-0-1}
  for the electron energy $E_{e}$.}
\end{figure}
Figure~\ref{Fig:eenlab-0-1} shows the effects of NNLO corrections 
for the observable $d\sigma / dE_e$. 
In Setup~1, the NNLO correction along the electron line
is negative, not exceeding the $-0.1\%$ level up to energies of
the order of $120$~GeV, where it shows a change of slope towards positive
values, reaching the level of $+1.5\%$ for the maximum available
energy. This feature reflects the behaviour of the NLO correction,
shown in the inset, which ranges from $+5\%$ to $-5\%$
in the region $m_e < E_e < 120$~GeV and 
drops to about $-18\%$ at the kinematical limit, signalling the presence
of enhanced IR effects because of missing available phase space for real
photon emission. For Setup~2, the IR enhancement of the NNLO corrections is
visible over the entire energy range, changing monotonically from about
$+0.5\%$ for $E_e \to m_e$ to about $3.2\%$ at the upper kinematical
limit. The NNLO correction along the muon line is similar in shape to the
one along the electron line but with reduced size by about a factor
of $30$. 

\begin{figure}[t]
\begin{center}
\includegraphics[width=0.95\textwidth]{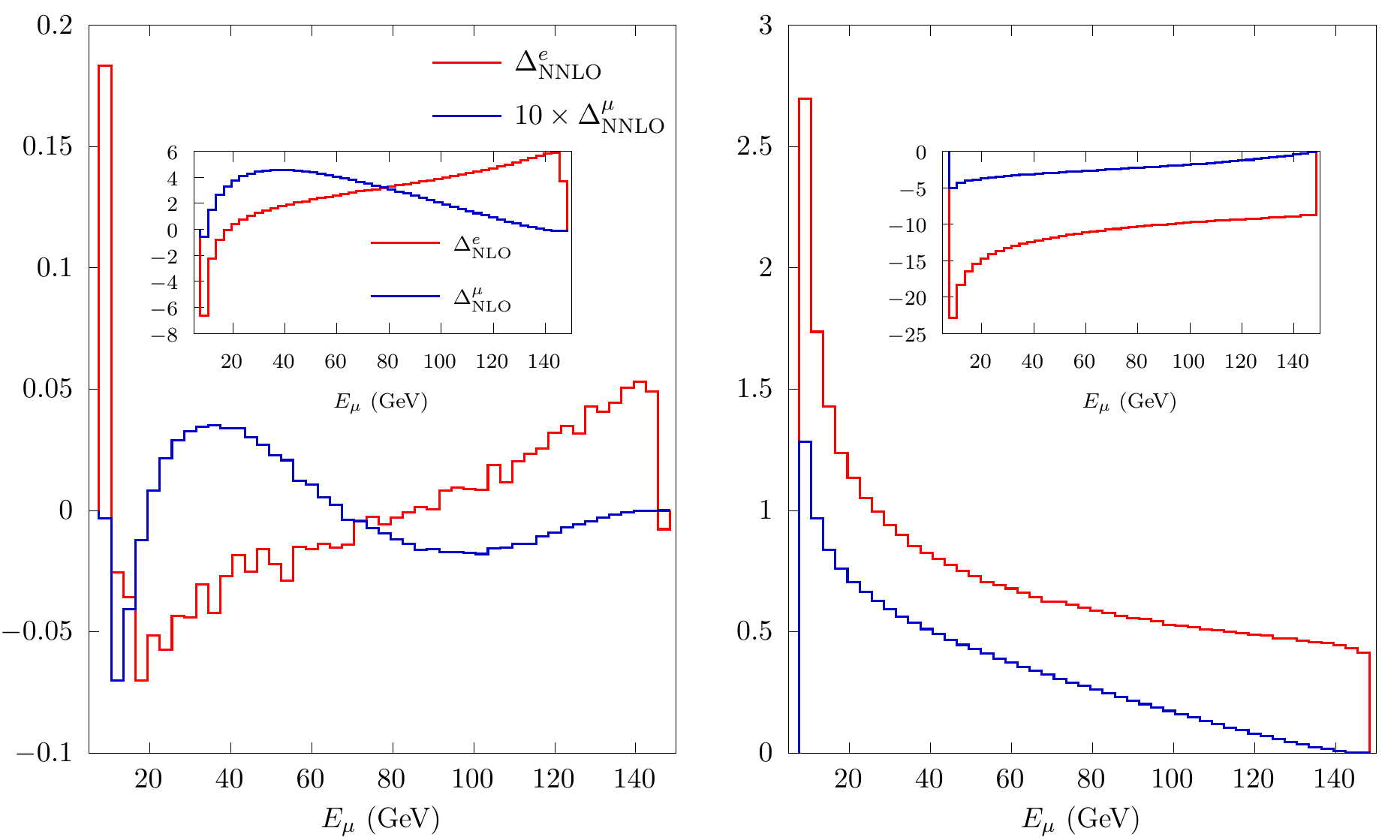}
\end{center}
\caption{\label{Fig:menlab-0-1} The same as Fig.~\ref{Fig:ethlab-0-1}
  for the muon energy $E_{\mu}$.}
\end{figure}
In Fig.~\ref{Fig:menlab-0-1} we show the effects of NNLO corrections on $d\sigma / dE_\mu$. 
In the case of Setup~1, the shapes of the contributions 
along the electron line (red histogram) and the muon line (blue histogram) 
are quite different. The former displays a sharp peak, of the order of
$+0.18\%$ in the first bin for the minimum muon energy and then ranges 
monotonically from $-0.07\%$ up to about $+0.05\%$.
The correction along the muon line shows
a more rich structure, never exceeding the size of $0.006\%$.
When including the acoplanarity cut (Setup~2),
the NNLO correction along the electron line becomes a monotonically
decreasing function of $E_\mu$, ranging from $+2.7\%$ to about $+0.4\%$.
The NNLO correction along the muon line (in blue) displays qualitatively
a similar behaviour. 

\begin{figure}[t]
\begin{center}
\includegraphics[width=0.95\textwidth]{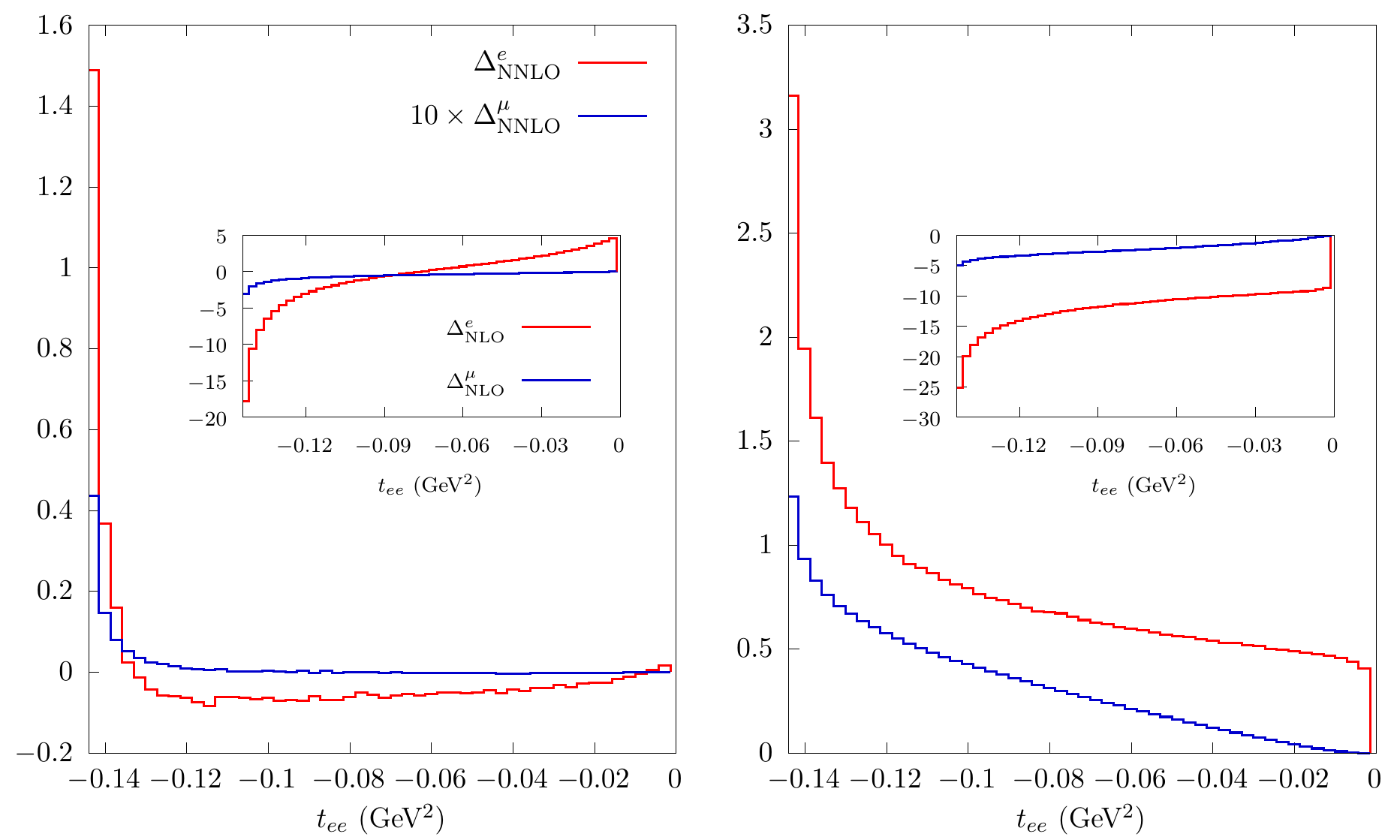}
\end{center}
\caption{\label{Fig:t24-0-1} The same as Fig.~\ref{Fig:ethlab-0-1}
  for the squared momentum transfer measured along the electron line, $t_{ee}$.}
\end{figure}
Figure~\ref{Fig:t24-0-1} shows the size of NNLO corrections on $d\sigma / dt_{ee}$. 
In Setup~1, the correction along the electron line (red histogram) 
is slightly negative, never exceeding the -0.1\% level, for
$-0.12\, \, \text{ GeV}^2 < t_{ee} < 0$~GeV$^2$. For comparison, it is
worth noticing that the NLO correction changes by about $10\%$, by looking
at red line in the inset. For lower values (larger absolute values)
of $t_{ee}$, the correction shows a steep increase to positive values,
reaching about $1.5\%$ at the kinematical
boundary, where the NLO correction reaches about $-18\%$ w.r.t. the
tree-level prediction. The NNLO correction along the muon line (blue histogram) 
displays a similar behaviour, as a function of $t_{ee}$, but with much
smaller size: it does not exceed the $10^{-4}$ level in the window
$-0.12\, \, \text{ GeV}^2 < t_{ee} < 0$~GeV$^2$ and reaches the size of
about $+0.04\%$ at the kinematical boundary.
The right panel of Fig.~\ref{Fig:t24-0-1} shows the NNLO corrections
when the acoplanarity cut is added (Setup~2). The electron-line
correction is positive and increasing with increasing $\vert t_{ee} \vert$,
from about $0.5\%$ up to $3.2\%$ at the boundary. This can be
expected since the corresponding NLO correction on the electron line
ranges from about $-10\%$ to $-25\%$, because of the increased importance
of the IR logarithm in the presence of cuts that tend to suppress
hard photon emission. Also the NNLO correction along the
muon line (blue histogram) is enhanced in the right panel ranging from
about $0\%$ up to $0.13\%$. Correspondingly, the NLO correction shown in
the inset ranges from $0\%$ to $-5\%$.
\begin{figure}[t]
\begin{center}
\includegraphics[width=0.95\textwidth]{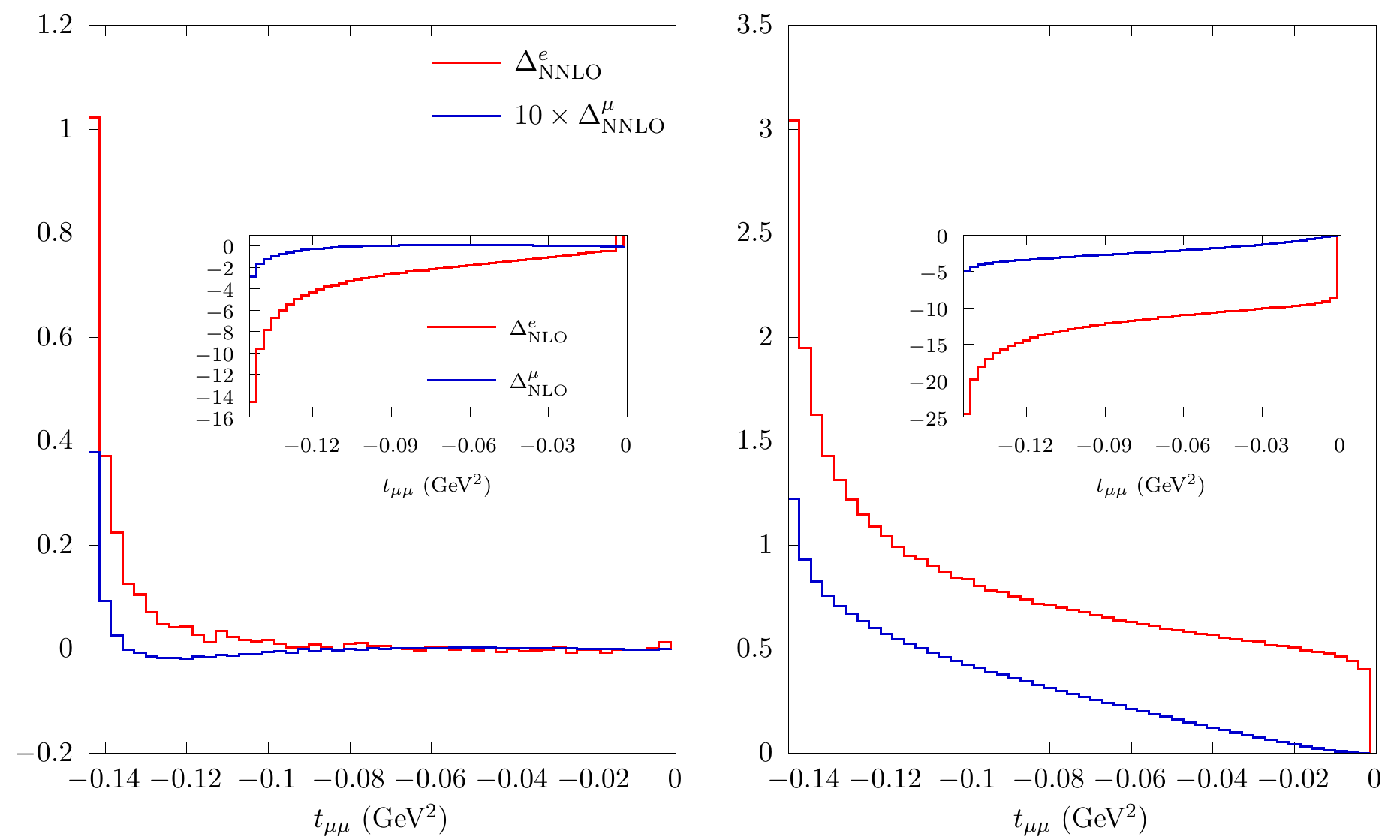}
\end{center}
\caption{\label{Fig:t13-0-1} The same as Fig.~\ref{Fig:ethlab-0-1}
  for the squared momentum transfer measured along the muon line, $t_{\mu\mu}$.}
\end{figure}

Finally, in Fig.~\ref{Fig:t13-0-1} we consider the size of NNLO
corrections on $d\sigma / dt_{\mu\mu}$. 
The behaviour of the corrections, as functions of $t_{\mu\mu}$, 
is similar to those
of Fig.~\ref{Fig:t24-0-1}.
In particular, for Setup~1, the correction along the
electron (muon) line ranges from about $0$\% to $1\%$~$(0.04\%)$ at
the kinematical boundary.
Also in the presence of the acoplanarity cut of Setup~2 (right panel),
the shapes of the corrections are similar to the ones of
Fig.~\ref{Fig:t24-0-1}, ranging from about $0.5\%$~$(0\%)$ to $3\%$ $(0.13\%)$
along the electron (muon) line. 

\subsection{Approximation to the full NNLO corrections}
\label{sec:numerics-full}
\begin{figure}[t]
\begin{center}
\includegraphics[width=0.95\textwidth]{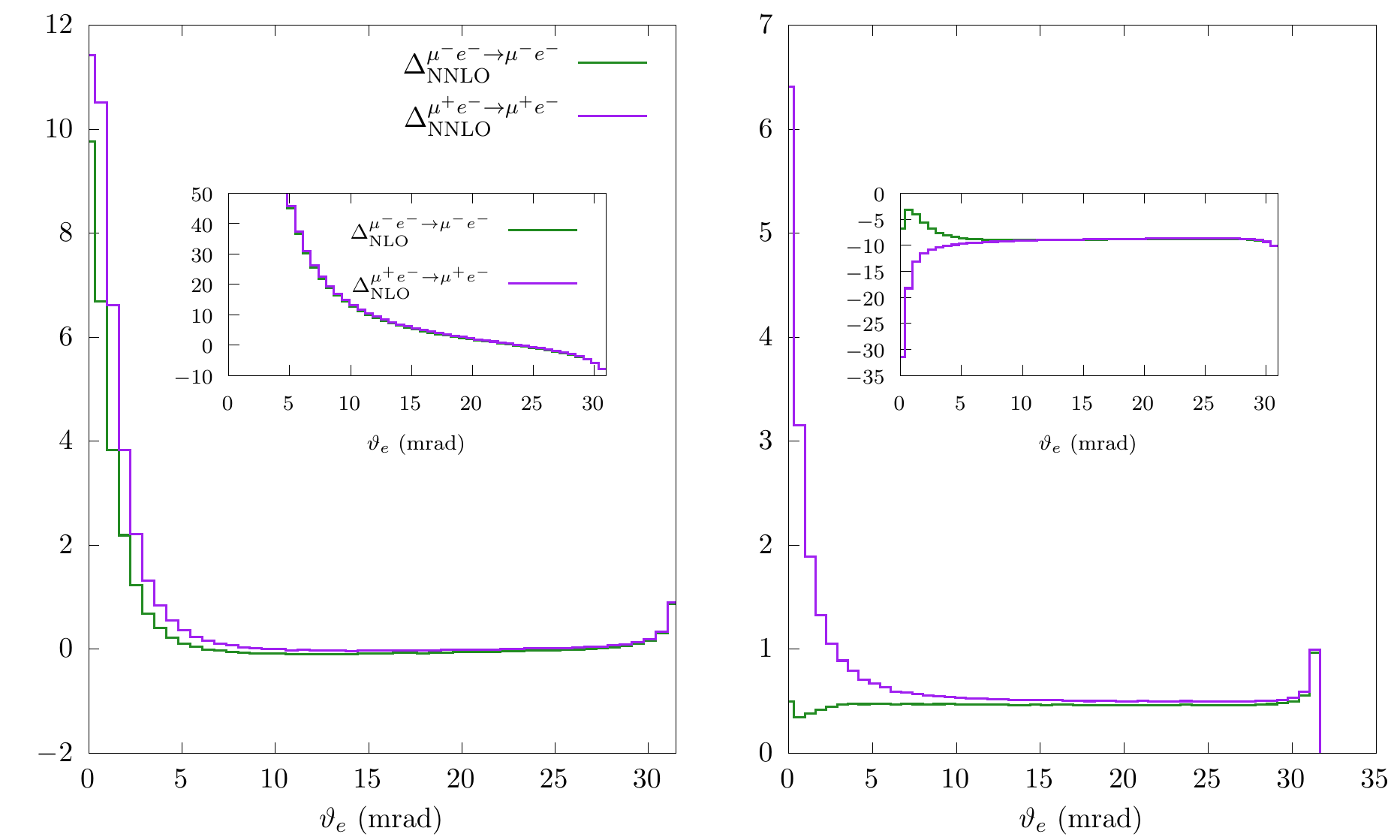}
\end{center}
\caption{\label{Fig:ethlab-int}
  Approximation to the full NNLO corrections, 
  according to Eq.~(\ref{eq:YFSapproxfull}) or (\ref{eq:YFSapproxfull2}), 
  as a function of $\vartheta_e$.
  The green histogram refers to the process $\mu^- e^- \to \mu^- e^- $ while 
  the violet histogram refers to $\mu^+ e^- \to \mu^+ e^- $. 
  Left panel: differential distribution in the presence of acceptance cuts
  only (Setup~1); right panel: the same as in the left panel
  with the additional acoplanarity cut of Setup~2.
  The insets report the NLO corrections, according to
  Eq.~(\ref{eq:delta_NLO-def}).}
\end{figure}
In Fig.~\ref{Fig:ethlab-int}, to be
compared to Fig.~\ref{Fig:ethlab-0-1}, we show how
the full (approximate) NNLO correction impacts the $\vartheta_e$
distribution. In absence of the acoplanarity cut (Setup~1, left
panel), we notice how the NNLO correction is similar to the one
stemming only from the electron line, except that at the left corner of the
distribution is a few $\%$ larger for the $\mu^+$ initiated process
and smaller of a similar amount when $\mu^-$ is in the initial
state. A more dramatic change is visible in Setup~2, when the
acoplanarity cut is applied. For the $\mu^+e^-\to\mu^+e^-$ case, the NNLO
corrections are similar to the electron-line corrections above $\sim 5\text{ mrad}$, but grow to $6.5\%$
as $\vartheta_e\to 0\text{ mrad}$. For the $\mu^-e^-\to\mu^-e^-$ case
instead, also in the region $\vartheta_e<5\text{ mrad}$ the NNLO correction
remains moderate at the $0.5\%$ level and almost flat. We remark that an
analogous difference between $\mu^+$ and $\mu^-$ is present also at
NLO (see the insets and Figs.~10 and 11 of Ref.~\cite{Alacevich:2018vez})
and that NNLO corrections tend to reduce the NLO
ones. The variation in the corrections for different charge of the muon can be ascribed to different radiation patterns,
which interfere destructively in one case and constructively in the
other one.

\begin{figure}[t]
\begin{center}
\includegraphics[width=0.95\textwidth]{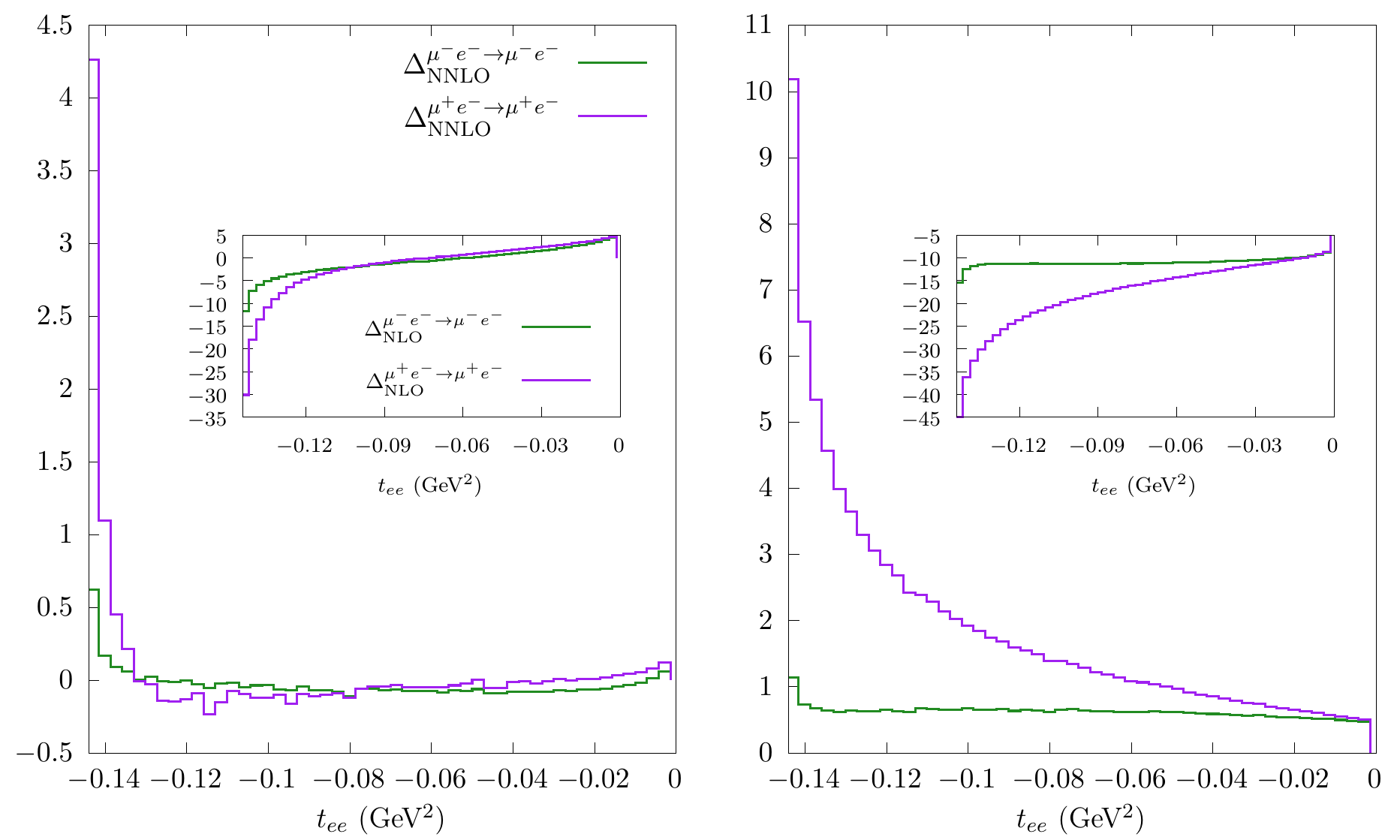}
\end{center}
\caption{\label{Fig:t24-int} The same as Fig.~\ref{Fig:ethlab-int}
  for the squared momentum transfer measured along the electron line, $t_{ee}$.}
\end{figure}
We finally show the NNLO effects on the $t_{ee}$ distribution
in Fig.~\ref{Fig:t24-int}, to be compared to
Fig.~\ref{Fig:t24-0-1}. Also in this case, for Setup~1, the largest
differences w.r.t. the corrections coming only from electron line occur
for $t_{ee}$ close to the kinematical limit
$t_{ee}^\text{min}\simeq-0.143\text{ GeV}^2$, where the phase space allowed for radiation is restricted
and IR effects are enhanced. For Setup~2, NNLO corrections are almost
flat and of order $1\%$ for incoming $\mu^-$, while increase in size going
from order $1\%$ to $10\%$ as $t_{ee}$ goes from $0\text{ GeV}^2$ to
$t_{ee}^\text{min}$ for incoming $\mu^+$.

\begin{figure}[t]
\begin{center}
\includegraphics[width=0.95\textwidth]{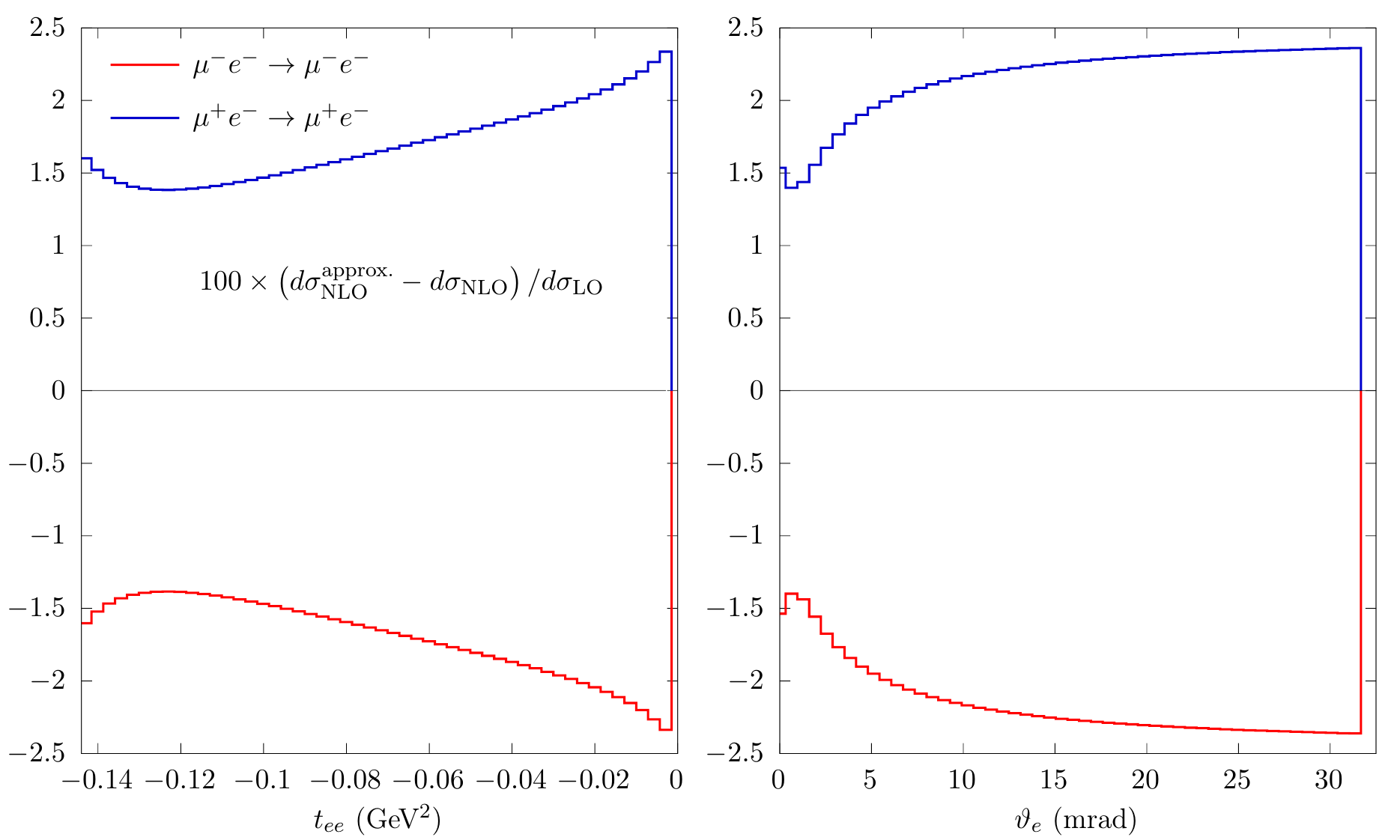}
\end{center}
\caption{\label{Fig:yfs-nlo}
Difference between NLO box diagrams approximated \emph{\`a la} YFS and the exact
calculation, as a function of the $t_{ee}$ (left) and $\vartheta_e$ (right) variables.
}
\end{figure}

After discussing the numerical impact of the approximate full two-loop
amplitude of Sect.~\ref{sec:nnlo_full-YFS} on relevant distributions,
we conclude by examining its theoretical accuracy, which is
necessarily an educated estimate since the exact calculation is not available
yet. Firstly, we show how the one-loop box diagrams are approximated by their IR
part as obtained from the YFS approach. In other terms, we compare the
exact NLO virtual amplitude $\mathcal{M}^{\alpha^1}$ of
Eq.~(\ref{eq:YFS2loop}) with the approximation
\begin{equation}
\widetilde{\mathcal{M}}^{\alpha^1} = \mathcal{M}^{\alpha^1}_e + \mathcal{M}^{\alpha^1}_\mu + Y_{e\mu}\mathcal{T},
\label{eq:YFSnlobox}
\end{equation}
which can be thought as the NLO version of
Eq.~(\ref{eq:YFSapproxfull2}), or, otherwise, the latter can be seen as the NNLO
generalization of the former. In
Fig.~\ref{Fig:yfs-nlo}, the difference of the exact NLO
differential cross section and the approximate one, relative to LO, is
shown as a function of the $t_{ee}$ (left) and $\vartheta_e$ (right) variables. We remark that the
difference $(d\sigma_\text{NLO}^\text{approx}-d\sigma_\text{NLO})$
depends on the charge of the incoming muon, as expected, but does not
depend on the acoplanarity cut because only the virtual amplitude is
modified. The approximation
differs from the exact result at the level of $1$-$2 \%$. We notice that
this is of the same order of
$\delta_\text{NLO}=\frac{\alpha}{\pi}\ln\left(m_\mu^2/m_e^2\right)\simeq 2.5\%$. This
simple observation brings us to estimate the accuracy of
Eq.~(\ref{eq:YFSapproxfull2}) for the full two-loop amplitude to be of the
order of
$\delta_\text{NLO}^2 =
\left(\frac{\alpha}{\pi}\right)^2\ln^2\left(m_\mu^2/m_e^2\right)\simeq
6\times 10^{-4}$.

\begin{figure}[t]
\begin{center}
\includegraphics[width=0.95\textwidth]{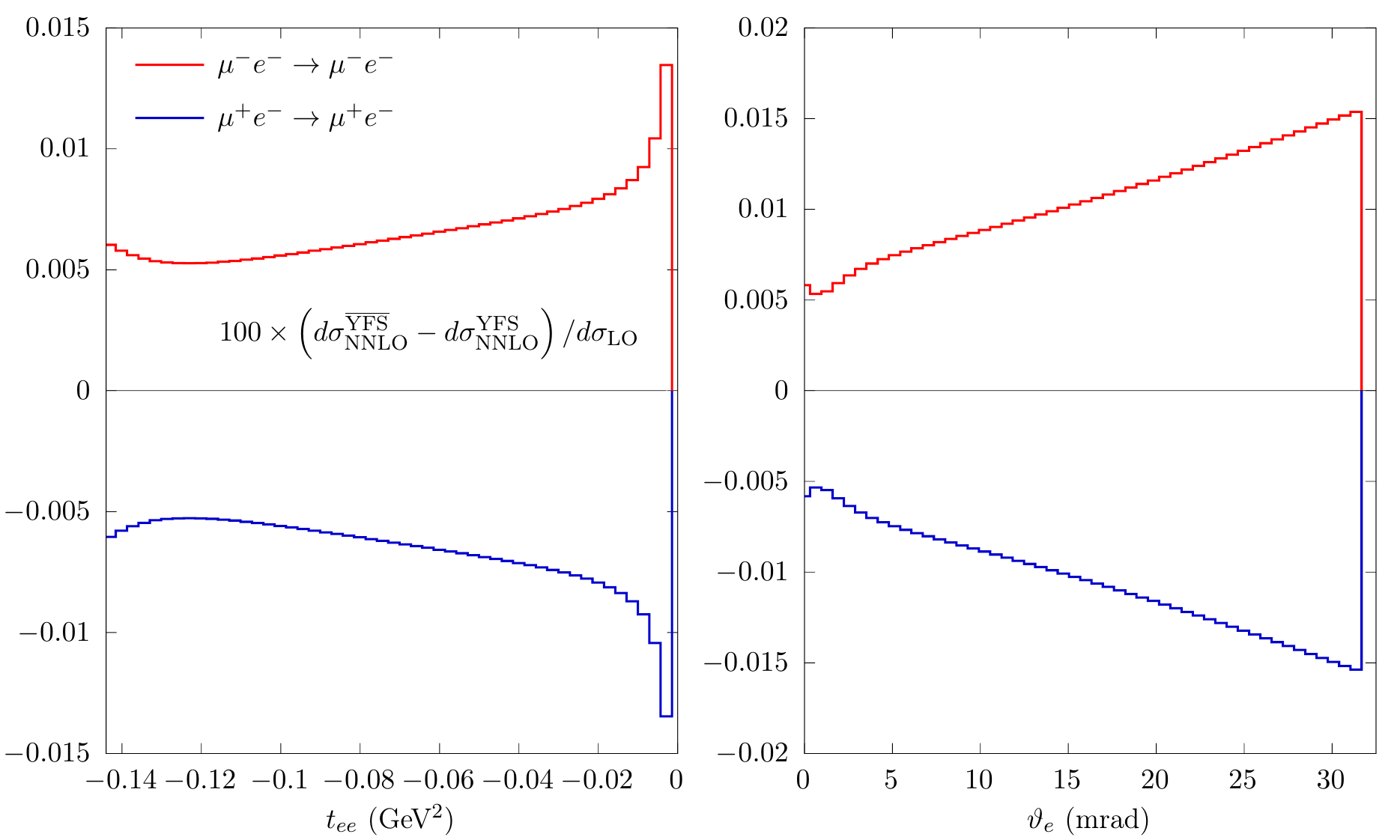}
\end{center}
\caption{\label{Fig:yfs-nnlo}
Difference between the alternative implementation of the YFS
approximation described at the end of Sect.~\ref{sec:numerics-full}
and the standard one, as a function of the $t_{ee}$ (left) and $\vartheta_e$ (right) variables.
}
\end{figure}

Our estimate is further corroborated by noticing that the IR
separation implied by Eq.~(\ref{eq:YFS2loop}) is not unique because
non IR-terms can be freely moved from the $Y$ factors to the
${\cal M}^{\alpha^{[1,2]},\text{R}}$ remnants. In our case, where we
necessarily set ${\cal M}^{\alpha^{2},\text{R}}=0$ for diagrams with at
least two photons connecting the $e$ and $\mu$ lines, we can for instance
switch off some non-IR terms in the second line of Eq.~(\ref{eq:YFSapproxfull2}) to probe
the size of the missing non-IR terms. For the sake of illustration,
we accomplish this by setting to zero non-IR terms (which are also not
singular in the limit of vanishing lepton masses) in the $C_0$ functions appearing in
Eq.~(\ref{eq:YFSfactors}). More explicitly, in the expression of the IR
$C_0$ function (see Eq.~(B.5) of Ref.~\cite{Dittmaier:2003bc})
\begin{align}
C_0  =
\frac{x_s}{m_1m_2\left(1-x_s^2\right)}&\left\{-\ln\left(\frac{\lambda^2}{m_1m_2}\right)\ln\left(x_s\right)-\frac{1}{2}\ln^2\left(x_s\right)
+ \frac{1}{2}\ln^2\left(\frac{m_1}{m_2}\right)
 \right.\nonumber\\
& \left. \ \ +\, 2\ln\left(x_s\right)\ln\left(1-x_s^2\right)  -
\frac{\pi^2}{6} + \text{Li}_2\left(x_s^2\right) \right.\nonumber\\
& \left. \ \ +\, \text{Li}_2\left(1-x_s\frac{m_1}{m_2}\right) + \text{Li}_2\left(1-x_s\frac{m_2}{m_1}\right)
\right\},
\label{eq:C0}
\end{align}
where $m_{1,2}=m_e\text{ or } m_\mu$, $\lambda$ is the vanishing
photon mass and $x_s$ is defined in Eq.~(B.7) of
Ref.~\cite{Dittmaier:2003bc}, we remove the finite terms in second and
third line of Eq.~(\ref{eq:C0}).
We label this further NNLO approximation as
$\overline{\text{YFS}}$ and in Fig.~\ref{Fig:yfs-nnlo} we compare it to the standard one at
differential level on the $t_{ee}$ and $\vartheta_e$
distributions. The plots show that the (IR-safe) differences lie below
the $2\times10^{-4}$ range (in
LO units), of the same size of the above estimate based on the NLO result.

\section{Summary and prospects}
\label{sec:summary}
In this work, we have analyzed the NNLO photonic 
corrections to $\mu^\pm e^- \to \mu^\pm e^-$ scattering and 
presented phenomenological results obtained with a related MC code,
named \textsc{Mesmer}. 
The study is motivated by the recent proposal of measuring the effective electromagnetic coupling constant in the space-like region by using this process
(MUonE experiment). From the measurement of the 
hadronic contribution to the running of $\alpha_\text{QED}$, the leading hadronic contribution to the muon anomaly 
can be derived according to an alternative approach to 
the standard time-like evaluation of $a_\mu^\text{HLO}$.

The calculation features the exact NNLO contribution of the
gauge-invariant subset of QED photonic corrections along the leptonic legs,
including all finite mass terms. The IR divergences are regularized by means
of a small photon mass and a slicing photon energy parameter to separate
soft from hard real radiation. The contribution of the vertex vacuum
polarization insertion has been removed, in order to avoid 
potentially large logarithmic contributions which are cancelled by
the real radiation of leptonic pairs, leaving the issue to a
future investigation. The complete NNLO calculation,
{\it i.e.} including also the up-down radiation contribution, has been
calculated in an approximate way, since the complete NNLO virtual amplitudes
are not yet available in the literature. Nevertheless, a YFS approach allows
to approximate the missing NNLO virtual amplitudes
with the correct inclusion of all IR enhanced terms. Moreover, the double-real
radiation matrix elements as well as the phase space are exact, 
including all finite lepton mass effects. Also the NLO corrections
to the $\mu^\pm e^-\to\mu^\pm e^-\gamma$ process, which are a 
part of the NNLO calculation, are exact. 

The developed formalism has been implemented in a fixed-order
generator able to reach NNLO accuracy for any
differential cross sections. The structure of the code is completely
general and the YFS approximation to the two-loop diagrams
with at least two virtual photons connecting the electron and muon lines 
can easily be replaced with the corresponding exact amplitude, once available. 

By means of the developed MC generator, we have extended a 
recent analysis from the NLO to the NNLO accuracy. 
In particular, we have considered a threshold electron energy
of 1~GeV in the laboratory frame and two basic reference event
selections: one involving only typical acceptance cuts of
the MUonE experimental detector for the final
state electron and muon; one including also a simplified acoplanarity
cut to restrict the events around the elasticity curve given by
the tree-level correlation between final state electron and muon
scattering angles. We have considered a number of differential distributions,
which allow to fully characterize the events of the MUonE experiment
and to outline the best data analysis strategy to extract the HLO
contribution to the running of the electromagnetic coupling constant
in the space-like region.

The phenomenological study of the different observables points out
that the size of the NNLO corrections, w.r.t. the LO differential cross-sections, is at the level of a few $10^{-4}$
for several regions of phase space in the presence of acceptance
cuts only, leaving some corners of phase space where the corrections
can grow up to the per cent level. As already remarked in the previous NLO
analysis, the only exception is the electron scattering angle distribution
which is particularly sensitive to photon radiation, in the region of
small scattering angles where the LO prediction tends to zero
as $\vartheta_e \to 0$. A similar feature of the LO prediction is also present 
for the muon scattering angle distribution,
but in a very small range around $\vartheta_\mu \to 0$. 
Moreover, the typical size of the NNLO corrections
grows up to the order of a few per mille in the presence of the acoplanarity
cut, reaching the several per cent level at the boundaries of the phase space.
This is a signal that IR terms start to dominate over the
rest of the matrix elements (as already noticed in the NLO analysis)
and definitely points to the relevance of the
next step of matching the fixed order
calculation with an all order exclusive resummation procedure
of multi-photon emission, for instance by generalizing to NNLO
accuracy the
matching of NLO corrections with a QED Parton Shower of
Ref.~\cite{Balossini:2006wc}. This step is left to future work and it is by
now under consideration. 

\acknowledgments
We are sincerely grateful to all our MUonE colleagues for stimulating
collaboration and many useful discussions, which are the framework of the
present study.
In particular we thank the PSI/Zurich group (P.~Banerjee, T.~Engel, A.~Signer
and Y.~Ulrich) for continuous exchange and careful cross-check of our
results, which helped to validate the technical precision of the developed
Monte Carlo program.
We are also indebted to P.~Mastrolia and A.~Primo for providing us
with updated expressions of the two-loop form factors.
The work of M.C. has been supported by the ``Investissements d'avenir, Labex
ENIGMASS''. 

\bibliographystyle{JHEP}
\bibliography{muone_twoloop}

\end{document}